\documentstyle[psfig]{mn}
\title{NGC~2401: A template of the Norma-Cygnus Arm's young population in the
Third Galactic Quadrant}
\author[Baume et al.]
{G. Baume$^{1,2}$, A. Moitinho$^{3}$, R.A. V\'azquez$^{1}$, G. Solivella$^{1}$,
 G. Carraro$^{2,4,5}$ and S. Villanova$^{2}$
\thanks{Based on observations collected at CTIO and ESO and CASLEO;
        Data is only available in electronic form at the CDS via
        anonymous ftp to {\tt cdsarc.u-strasbg.fr (130.79.128.5)} or
        via {\tt http://cdsweb.u-strasbg.fr/cgi-bin/qcat?J/A+A//};
        email:gbaume@fcaglp.unlp.edu.ar(GB)}\\
$^1$Facultad de Ciencias Astron\'omicas y Geof\'{\i}sicas de la
    UNLP, IALP-CONICET, Paseo del Bosque s/n, La Plata, Argentina\\
$^2$Dipartimento di Astronomia, Universit\`a di Padova,
    Vicolo Osservatorio 2, I-35122 Padova, Italy\\
$^3$CAAUL, Observat\'orio Astron\'omico de Lisboa, Tapada da Ajuda,
    1349-018 Lisboa, Portugal\\
$^4$Astronomy Department, Yale University, P.O. Box 208101,
    New Haven, CT 06520-8101 USA\\
$^5$Departamento de Astronom\'{\i}a, Universidad de Chile,
    Casilla 36-D, Santiago, Chile
    }

\date{\it Received **; accepted **}
\pubyear{2005}

\begin{document}
\maketitle
\title{NGC~2401}

\begin{abstract}
     Based on a deep optical CCD ($UBV(RI)_C$) photometric survey and on the Two-Micron
     All-Sky-Survey (2MASS) data we derived the main parameters of the
     open cluster NGC~2401. We found this cluster is placed at 6.3 $\pm$
     0.5 kpc ($V_O - M_V = 14.0 \pm 0.2$) from the Sun and is 25~Myr old, what
     allows us to identify NGC~2401 as a member of the young population belonging
     to the innermost side of the extension of the Norma-Cygnus spiral--arm in the
     Third Galactic Quadrant. A spectroscopic study of the emission star
     LSS~440 that lies in the cluster area revealed it is a B0Ve star; however,
     we could not confirm it is a cluster member. We also constructed the cluster
     luminosity function (LF) down to $V \sim 22$ and the cluster initial mass
     function (IMF) for all stars with masses above $M \sim 1-2 M_{\sun}$. It was
     found that the slope of the cluster IMF is $x \approx 1.8 \pm 0.2$. The
     presence of a probable PMS star population associated to the cluster is
     weakly revealed.
\end{abstract}


%

\section{Introduction}

The open cluster NGC~2401 (=OCL 588 = C0727-138) is an almost
unstudied compact grouping of faint stars, but for identification
and eye estimates of its angular size and richness. According to
the Lyng\aa\ (1987) classification, this object is a Trumpler
class II 3 p with about $2^{\prime}$ in diameter, which is the
same diameter listed in Dias et al. (2002). A recent
photometric study on this area carried out by Sujatha et al.
(2004) yielded an intriguing result where a relatively far cluster
(located at a distance of about 3.1 kpc) shows only a color excess
$E_{B-V} = 0$. This fact alone deserves our total attention.
Moreover, since NGC~2401 is placed in the Puppis region ($l =
229.67^{\circ}$; $b = +1.85^{\circ}$) in the poorly studied
Third Quadrant of the Galaxy, the determination of its basic
parameters will contribute, together with other cluster studies in
the region, to describe more precisely the spiral structure and
the star formation history in this part of the Galaxy. This region
contains several distant clusters through which information on the
kinematics and evolutionary status of the stellar population in
the outermost parts of the galactic disk can be obtained. In fact,
since the detection of the Canis Major over-density (see p.e.
Martin et al. 2004, Momany et al. 2004 for more details), there is
a growing interest to get a better description of the stellar
population in this region of the Galaxy.

Preliminary results of NGC~2401 can be found in a CCD $UBVRI$
photometric database by Moitinho (2001, 2002) containing brief
information for 30 open clusters in the galactic longitude range
$217^{\circ}< l < 260^{\circ}$. Also recently, Giorgi et al. (2002, 2005),
Carraro \& Munari (2004), Baume et al. (2004), Carraro et al. (2005a) and
Moitinho et al. (2005) published studies of a series of largely overlooked
open clusters in the same galactic region.

In this paper we present a detailed study of the membership,
reddening, distance and age of NGC 2401. They were derived
together with the Luminosity and Initial Mass Functions, two
distributions which are fundamental tools for understanding the
star formation mechanisms and related astrophysical problems. We
also present the first spectrum analysis of the peculiar emission
star LSS~440 located in the cluster field and try to clarify the
possible connection of both objects.

In the next section we introduce the observational material.
Sect.~\ref{sec:analysis} is aimed at analyzing the photometric diagrams
and to determine memberships. In Section~\ref{sec:lss} we discuss the LSS~440
main properties; in Sect.~\ref{sec:LFIMF} we construct and discuss both, the LF
and IMF. In Section~\ref{sec:PMS} we studied the possible presence of pre-main
sequence (PMS) stars and, finally, in Sects.~\ref{sec:disc} and \ref{sec:conc} we
discuss the main findings of this investigation and present the
conclusions respectively.

\begin{figure}
\begin{center}
\centerline{\psfig{file=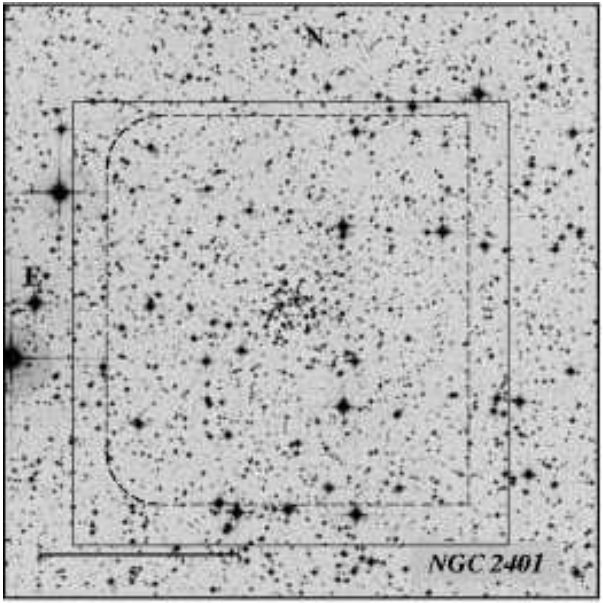,width=8cm}}
\caption{Second generation Digitized Sky Survey (DSS-2), red filter image of the
field of NGC~2401. The areas covered by the CTIO and ESO observations are
indicated by solid and dashed lines, respectively.}
\label{fig:dss}
\end{center}
\end{figure}

\section{Data set}

\subsection{Optical photometric data} \label{sec:data1}
CCD $UBV(RI)_C$ images of NGC~2401 were acquired with the CTIO
0.9m telescope during an observing run in January 1998.
Observations, reductions, error analysis, and comparison with
other photometries were thoroughly described in Moitinho (2001).
Additionally, further CCD $V(I)_C$ images were obtained for this
object at ESO - La Silla - with the EMMI camera mounted on NTT in
the night of December 9, 2002. Typical seeing was about $1''$. The
camera has a mosaic of two $2048 \times 4096$ pixels CCDs which
samples a $9\farcm9 \times 9\farcm1$ field. The images were binned
$2 \times 2$, resulting a scale of $0\farcs332$/pix.
Details on the reductions of these kind of data were given in
Baume et al. (2004). The fields covered by CTIO and ESO observations
are shown in Fig~\ref{fig:dss}.

Combining both data sets required a detailed comparison of their
plate scales and photometric scales. While matching the X,Y pixel
coordinates from both sets, we noticed that no one-to-one
correspondence could be satisfactorily achieved through a simple
linear transformation. Indeed, the residuals of the transformation
displayed a jump in the middle of the NTT X axis. This is likely
due to the junction between the two CCDs that compose the EMMI
camera. Once taking this effect into account, the NTT coordinates
were transformed without trouble to the system of the CTIO X-Y
positions, and sources from both sets were cross-identified. The
RMS residuals of the transformations were around 0.15 CTIO pixels
(1 CTIO pixel $\approx 0\farcs39$). The NTT $VI$ measurements
were then tied to the CTIO photometry through linear
transformations of the form $V_{CTIO} - V_{NTT} = \alpha_0+
\alpha_1 (V-I)_{NTT}$; $(V-I)_{CTIO} =\beta_0 + \beta_1
(V-I)_{NTT}$, using stars with estimated errors less than 0.015
mag, which essentially corresponds $V_{CTIO} < 18$mag. The rms of
the photometric transformations were $\approx 0.02$ mag. Both
photometries were then combined by averaging the measurements
weighted by their errors. The errors in the CTIO magnitudes are
described in Moitinho (2001), and are taken to be the dispersion
of the measurements when repeated observations were available, or
the errors output by ALLSTAR in the case of single measurements.
For the NTT magnitudes the adopted errors were those given by
ALLSTAR.

\begin{figure}
\begin{center}
\centerline{\psfig{file=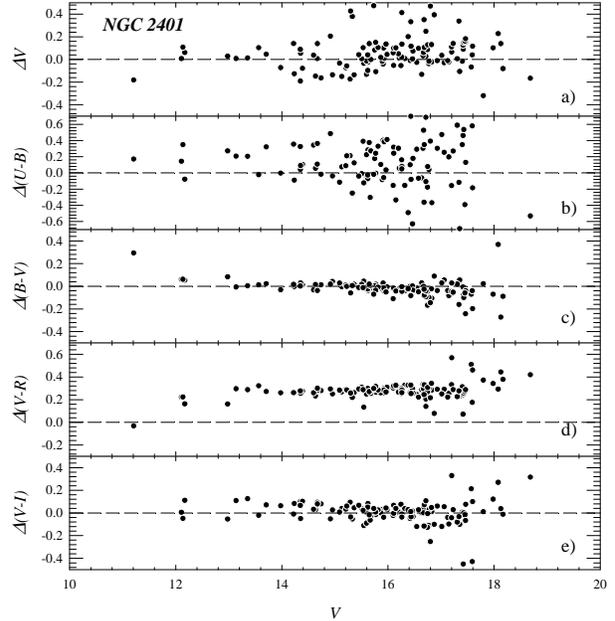,width=8cm}}
\caption{Comparison of our photometry with Sujatha et al. (2004) in the sense
"Our data - Their data" (115 stars). Mean differences and their standard
deviations (sd) were:
$\Delta V = +0.05 \pm 0.14(sd)$;
$\Delta (B-V) = -0.02 \pm 0.10(sd)$;
$\Delta (U-B) = +0.06 \pm 0.39(sd)$;
$\Delta (V-R) = +0.28 \pm 0.07(sd)$ and
$\Delta (V-I) = +0.00 \pm 0.11(sd)$}
\label{fig:deltas}
\end{center}
\end{figure}

We also cross correlated our data with the ones given by Sujatha
et al. (2004) to compare both photometric sets. Figure
~\ref{fig:deltas} shows a relatively good agreement in the mean
values for $V$ magnitudes and $B-V$ and $V-I$ colors, though the $V$
magnitude spread is quite significant. In addition, there is a
noticeable shift between both data sources when $U-B$ and $V-R$
colors are compared. In particular, the spread in $U-B$ is simply
huge. These differences could be explained by the different spatial
resolution of each observation set (some crowded areas in Sujatha et al.
data were resolved by us), zero point differences in their $U-B$ and
$V-R$ measures probably caused by poor atmospheric conditions in the
Sujatha et al. observing run and the rather poor quality of their $U$
filter. This way, the strange color excess value $E_{B-V} = 0$ computed by
them can be explained. It is worth mentioning that our data produce coherent
fitting solutions over all the photometric diagrams (see in advance
Sect.~\ref{sec:analysis} and Fig.~\ref{fig:cmd1}). On the other hand,
our photometric data are based on a more general and homogeneous
data set (Moitinho 2001), which are in agreement to previous studies
(when the comparison was possible). Therefore, we can conclude our data
are right at all.

\subsection{Spectroscopic data} \label{sec:data2}
Spectral data for the star LSS 440 ($\alpha_{2000} =
07:29:30.1$; $\delta_{2000} = -13:59:13$) were obtained at the
215-cm telescope of CASLEO (Argentina) during the nights of
December 6, 7 and 9, 2004. Observations were carried out with a
REOSC-DS Cassegrain spectrograph equipped with a Tek 1024 $\times$
1024 detector. The grating was successively centered at
$7^{\circ}~15^{\prime}$, $8^{\circ}~30^{\prime}$ and
$11^{\circ}~30^{\prime}$ (on different nights) to get a full
spectral range coverage from 3400 \AA~to 6750 \AA~in wavelength.
The dispersion was 2.5 \AA/pixel (resolution $\approx 1800$). Two
30 minutes exposure time spectra were taken in each grating
position to remove undesirable events such as cosmic rays and to
improve the signal--noise ratio of the final spectrum. Comparison
lamp (HeNeAr) spectra were acquired between each pair of
exposures. Spectra were reduced using IRAF\footnote{IRAF is
distributed by NOAO, which are operated by AURA under cooperative
agreement with the NSF.} routines like IMRED, CCDPROC, TWODSPEC
and ONEDSPEC using the typical procedure.

\subsection{Infrared data and astrometry} \label{sec:data3}

Available catalogues, such as 2MASS, are of fundamental
importance to perform a more complete analysis of any sky region.
Using the CCD X--Y positions of our data we computed their
equatorial coordinates. First of all, a matched list of X--Y and
RA, DEC was built by visually identifying about 20-30 TYCHO-2 (H{\o}g
et al. 2000) and 2MASS stars in each cluster field. The stars in
the list were used to obtain transformation equations to get
equatorial coordinates for all our measured stars. In a second
step we use a computer routine to cross-identify all our sources
in common with the same catalogues by matching the equatorial
coordinates to the catalogued ones. The rms of the residuals were
$\sim 0\farcs15$, which is about the astrometric precision of the
2MASS catalogue ($\sim 0\farcs12$), as expected since most of the
coordinates were retrieved from this catalogue.

This procedure allowed to build a photometric $UBVRIJHK$
catalogue that constitutes the main observational database used in
this study. Table 1 includes our X--Y positions (in
arcseconds, see Fig.~\ref{fig:xy}), equatorial coordinates (epoch
2000.0), optical and infrared photometry. This table is only
available electronically.

\section{Analysis} \label{sec:analysis}

\subsection{Cluster center and size} \label{sec:size}

\begin{figure}
\centerline{\psfig{file=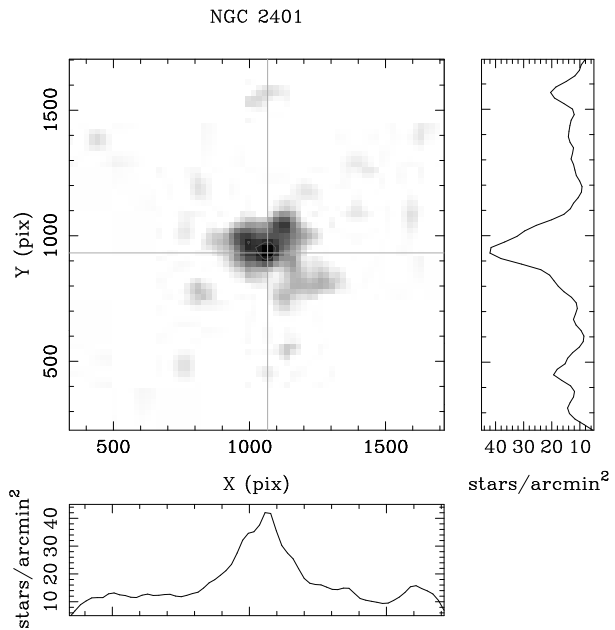,width=8cm}}
\caption{First approximation to the center determination for NGC~2401. Top Left:
density map of the selected stars in the field. Top right: Y cut through the
density maximum. Bottom: X cut through the density maximum.}
\label{fig:center}
\end{figure}

As a first step, we estimated the position of the cluster center.
This procedure was done by a combination of a visual
inspection of the DSS-2 plates and the method given by Moitinho
(1997). In this later case, the surface stellar distribution was
convolved with a kernel and the center is adopted as the point of
maximum density. In this case a $100 \times 100$ CTIO pixels
($\approx 0\farcm66 \times 0\farcm66$) gaussian kernel was
used to smooth out the details of the spatial distribution and
give a precise idea of the center position.
Fig.~\ref{fig:center} illustrates the method. The derived center
was $\alpha_{2000} = 07:29:25.2$; $\delta_{2000} = -13:57:54$,
nearby to the one given by Dias et al. (2002) or by the $SIMBAD$
database.

\begin{figure}
\begin{center}
\centerline{\psfig{file=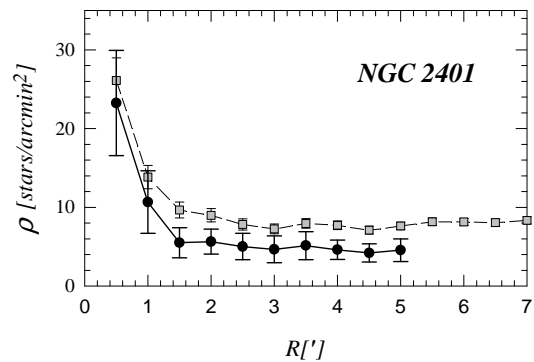,width=7cm}}
\caption{Radial density profiles for NGC~2401. Filled circles: CCD data. Grey
squares: 2MASS data. Poisson error bars ($1\sigma$) are also shown.}
\label{fig:radial}
\end{center}
\end{figure}

The second step was to compute the cluster radial density
profile by counting stars in a number of successive rings, $0\farcm5$
wide, and dividing the counts by the correspondent ring's area. We
applied this method to both, the optical and the 2MASS infrared
data. The respective radial density profiles are shown in
Fig.~\ref{fig:radial}. We appreciate a well shaped radial stellar
density profile for the cluster with an important central
concentration up to $\sim 1\farcm5$ but we adopted a radius of
$R = 2\farcm5$ since at this value the stellar density reaches
the (residual) field density. Our adopted value is a bit larger
than the diameter of about $2^{\prime}$ given by Lyng\aa~(1987)
and Dias et al. (2002). It seems that these works refer to the
object size as only the very central part of the cluster. On the
other hand, the radius we computed here is generous enough to
include cluster stars (mainly faint ones) that could be placed a
bit far from the cluster center. In any case, our data completely
covers the cluster together with an important part of its
surrounding field.

\subsection{Optical data and cluster membership} \label{sec:member}

\begin{figure}
\begin{center}
\centerline{\psfig{file=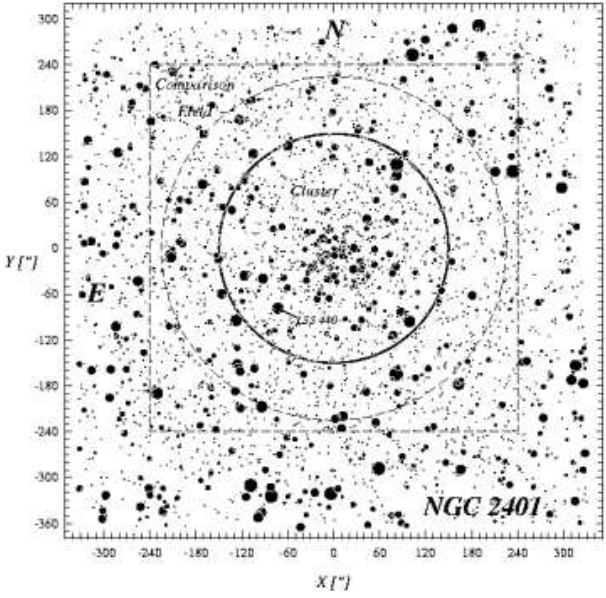,width=8cm}}
\caption{Finding chart of the observed region in NGC~2401 ($V$ filter). The black
solid circles, $2\farcm5$ in radius indicates the adopted angular size for the
cluster (see Sect.~\ref{sec:size} and Fig.~\ref{fig:radial}). The short--dashed
lines indicate the area adopted as a {\it Comparison Field} ($CF$).
For a coordinate reference, the $X=0$; $Y=0$ values correspond to the cluster
center coordinates $\alpha_{2000}=07:29:25.2$; $\delta_{2000}=-13:57:54$ and
all $X-Y$ are expressed in arcseconds.}
\label{fig:xy}
\end{center}
\end{figure}

\begin{figure*}
\centering
\centerline{\psfig{file=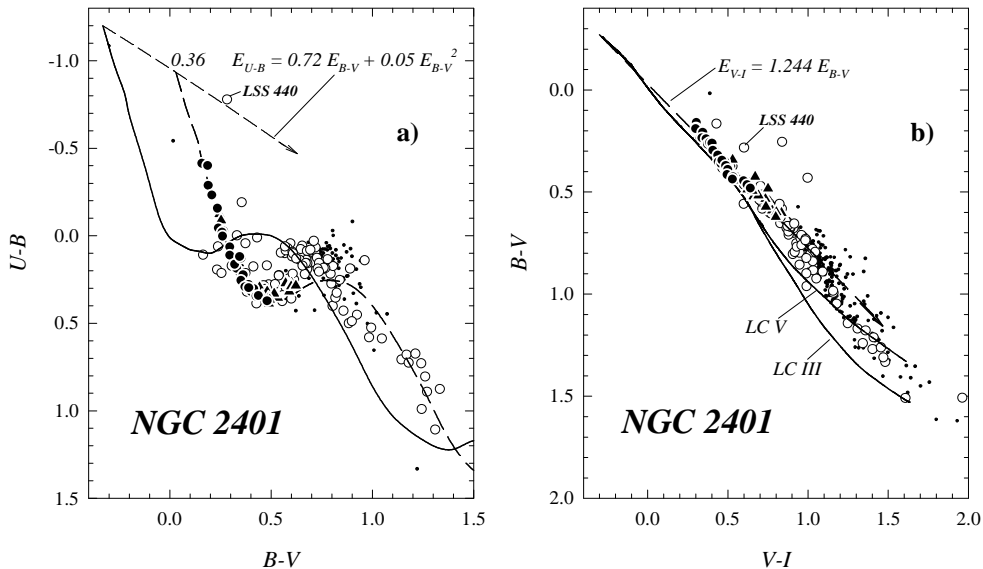,width=13cm}}
\caption{Optical TCDs of stars located inside the adopted radius
of NGC~2401 {\bf a)} $U-B$ vs. $B-V$ diagram. The symbols have the following
meaning: black circles are adopted likely member stars ($lm$), black triangles
are probable member stars ($pm$), white circles are non-member stars ($nm$), and
dots are stars without any membership assignment. The solid line is the
Schmidt-Kaler (1982) ZAMS, whereas the dashed one is the same ZAMS, but shifted
by the adopted color excess (see Sect.~\ref{sec:parameters}). The dashed arrow
indicates the normal reddening path. {\bf b)} $B-V$ vs. $V-I$ diagram. Symbols
and lines have the same meaning as in panel a).}
\label{fig:ccd1}
\end{figure*}

Information about proper motions and radial velocities
for the stars in the area of NGC~2401 is scarce.
The UCAC2 catalogue (Zacharias et al. 2004) provides proper motion
data for some stars in these area down to a magnitude about $V = 16$
that could help to perform a membership assignment. However, uncertainties
of these measurements ($\sim 10~mas/yr$) avoid using proper motions in a
efficient way. Therefore, the only way to obtain a reliable membership
assignment is by analyzing the individual position of the stars onto the
several photometric diagrams (e.g. Baume et al. 2003, 2004).

\begin{figure*}
\centering
\centerline{\psfig{file=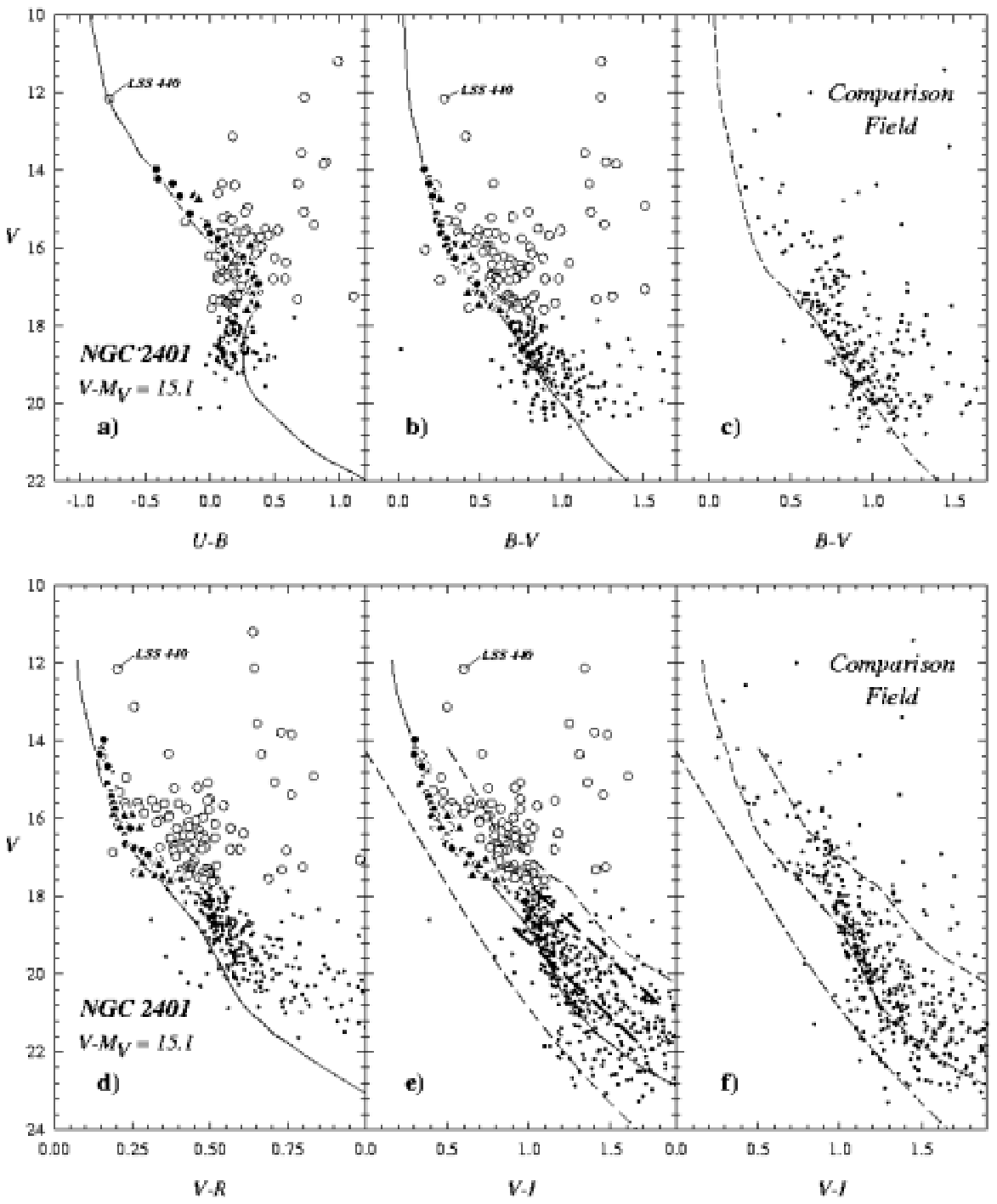,width=13cm}}
\caption{Optical CMDs of stars located inside the adopted radius of NGC~2401
($'cluster~area'$, panels a, b, d and e) and in the adopted $CF$ (panels c and f).
Symbols have the same meaning as in Fig.~\ref{fig:ccd1}. The solid lines are the
Schmidt-Kaler (1982) and Cousins (1978ab) empirical ZAMS and MS respectively,
corrected for the effects of reddening and distance. The adopted apparent
distance modulus is $V-M_{V} = 15.1$ ($V-M_{V} = V_{0}-M_{V} + 3.1 E_{B-V}$, see
Sect.~\ref{sec:parameters}). Dashed lines showed on the $CF$ diagrams
have the same meaning as the curves in the other panels: ZAMS, MS and the
adopted envelopes used to compute the LF (see Sect.~\ref{sec:LFIMF}). See
Sect.~\ref{sec:PMS} for the meaning of long dashed right lines on panel e)}
\label{fig:cmd1}
\end{figure*}

The optical Two Color Diagrams (TCDs) and Color Magnitude Diagrams
(CMDs) of NGC~2401 are shown in Figs.~\ref{fig:ccd1} and
\ref{fig:cmd1} respectively. The TCDs of Figs.~\ref{fig:ccd1}, and
the CMDs of Figs.~\ref{fig:cmd1}ab and \ref{fig:cmd1}de include
all the stars inside the adopted radius for the cluster (see
Sect.~\ref{sec:size}). For reference purpose, we adopted a
$Comparison~Field$ ($CF$) defined by the short--dashed lines in
the finding chart of Fig.~\ref{fig:xy}. This $CF$ was chosen in
such a way that its area equals the corresponding to the cluster
and it was also totally covered by the NTT observations. The CMDs
of the $CF$ are shown in Figs.~\ref{fig:cmd1}c and
\ref{fig:cmd1}f.

Different CMDs show different limiting $V$ magnitude going
from $\sim 19$ in $U-B$ to $\sim 23$ in $V-I$ due to the not
similar wavelength sensitivity of the detectors, on a side, and
also because we combined data from two sources (see
Sect.~\ref{sec:data1}), on the other. Naturally, deep $V-I$ data
came from the NTT.

All the diagrams confirm that NGC~2401 presents a very
sharp, clear and blue main sequence (MS) above $V \sim 17-17.5$.
For membership assignments, individual positions down to the
magnitude limit were examined in all the photometric diagrams
using the procedure described in Baume et al. (2004). This is:

\begin{itemize}
\item If stars brighter than $V \sim 16.5$ have coherent locations in all the TCDs
and CMDs along the MS, they were adopted as likely members ($lm$). We shall
discuss below the case of LSS 440 (see Sect.~\ref{sec:lss}).

\item Dimmer stars with magnitudes in the range $V \sim 16.5-17.5$ in the same
conditions were considered only as probable members ($pm$).

\item If some stars are brighter than $V \sim 16.5$, well placed onto the TCDs of
Fig.~\ref{fig:ccd1} but located a bit over the ZAMS on Fig.~\ref{fig:cmd1}, they
were still considered as $pm$ since their magnitude offsets could be due to a
probable binarity effect. This was undoubtedly the case of stars $\#~66$ and
$\#~75$ which, unlike likely members, appeared above the MS ($\sim 0.5$
mag).

\item Finally, the number of likely and probable member stars in each magnitude
bin must keep a reasonable agreement with the counts that were obtained when $CF$
stars were adequately subtracted from the $'cluster~area'$
(see Sect.~\ref{sec:LFIMF}).
\end{itemize}

At fainter magnitudes, contamination by field stars becomes
severe, preventing an easy identification of faint cluster
members. Anyway, it was still possible to determine the number of
probable faint cluster members in a statistical way (see
Sect.~\ref{sec:LFIMF} and Sect.~\ref{sec:PMS}).

\subsection{Cluster parameters} \label{sec:parameters}
\subsubsection{Optical data}

\begin{figure*}
\centering
\centerline{\psfig{file=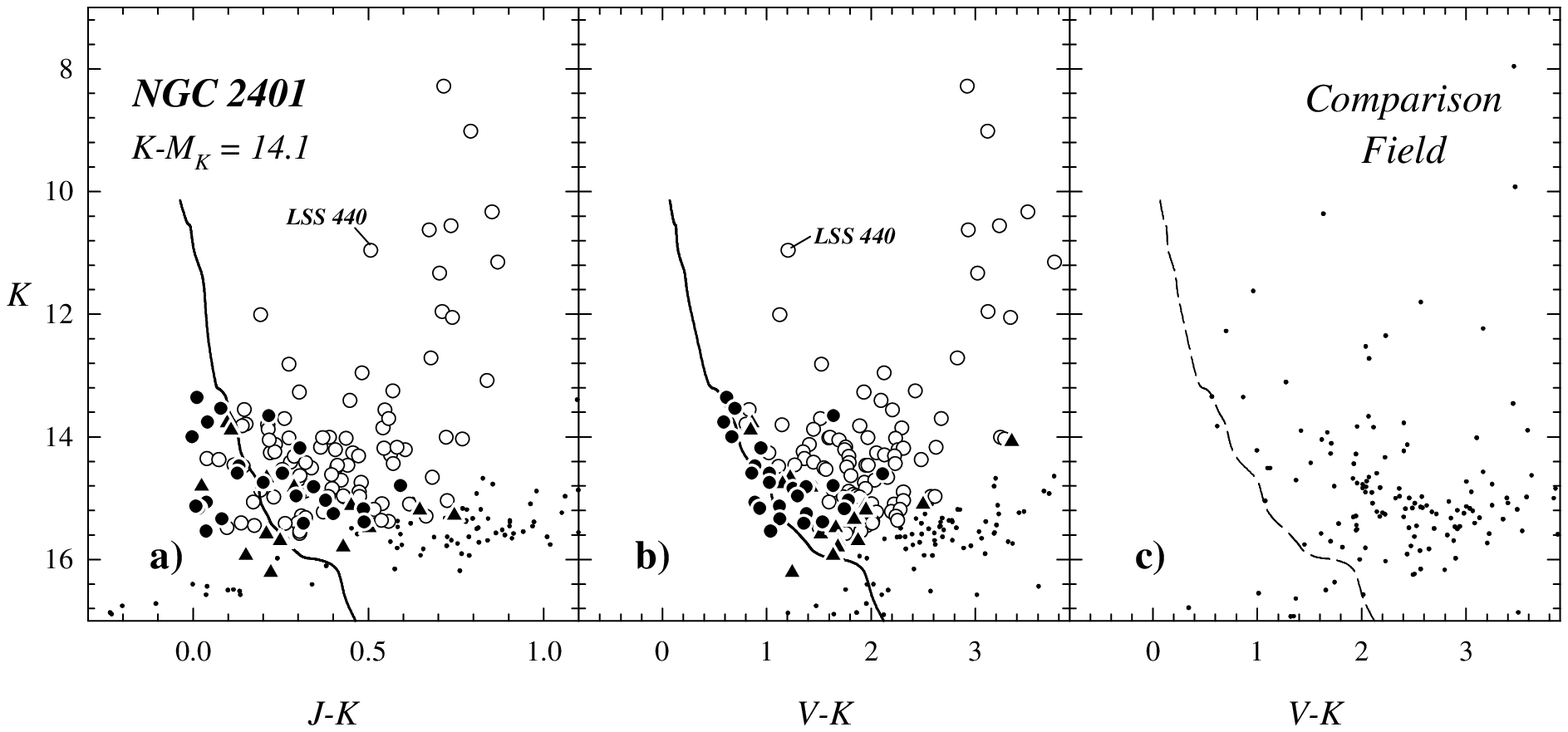,width=13cm}}
\caption{Infrared CMDs of stars located inside the adopted
radius of NGC~2401 (panels a and b) and in the adopted comparison field (panel c).
Symbols have the same meaning as in Fig.~\ref{fig:ccd1}. The solid lines are the
Koornneef (1983) empirical MS, corrected for the effects of reddening and distance
($K-M_{K} = 14.1 = V_{0}-M_{V} + (3.1 - 2.78)~E_{B-V}$, see
Sect.~\ref{sec:parameters}). Dashed line showed on the comparison field has the
same meaning as the curves in the other panels.}
\label{fig:cmd2}
\end{figure*}

This region of the Third Galactic Quadrant is characterized by a
reddening slope and a ratio of the total to selective absorption
($R = A_V / E_{B-V}$) that can be considered normal (Moitinho
2001). Since the picture presented in the $B-V$ vs. $V-I$
diagram in Fig.~\ref{fig:ccd1}b agrees with this concept, we
adopted standard ratios $E_{U-B} / E_{B-V} = 0.72 + 0.05~E_{B-V}$
and $E_{V-I} / E_{B-V} = 1.244$ (Dean et al. 1978), which implies
$R = 3.1$, to shift the Schmidt-Kaler (1982) ZAMS, the intrinsic
lines from Cousins (1978ab) and the Girardi et al. (2000)
isochrones in the TCDs and CMDs. By restricting to $lm$
and $pm$ stars, we found mean excess values and standard
deviations $E_{B-V} = 0.36 \pm 0.01$ and $E_{U-B} = 0.27 \pm
0.01$. As for the cluster distance modulus we found $V-M_{V} =
15.1 \pm 0.2$ (error from eye-inspection) by the fitting method.
This distance modulus combined with a mean visual absorption
$A_V=1.12$ places NGC~2401 at $6.3 \pm 0.5$~kpc from the Sun.

The age of the cluster was derived superposing the isochrones of
Girardi et al. (2000), computed with mass loss and overshooting
(see Fig.~\ref{fig:age}) and solar metallicity, and looking for
those ones that produce the best fit over the stars along the upper
MS. This method yields that NGC~2401 is about $20 \pm 5$ Myr old.
Another age indicator comes from the inferred earliest spectral
type at the upper MS of the cluster. In NGC 2401, the earliest MS
star may have a spectral type B3 and, according to the calibration
given by Meynet et al. (1993), the corresponding age is about 30
Myr. We adopted then $25 \pm 5$ Myr as a good estimation for
NGC~2401 age.

\subsubsection{Infrared data} \label{sec:ir}

As a control of the optical findings, we built up the CMDs
indicated in Fig.~\ref{fig:cmd2} using 2MASS data (see
Sect.~\ref{sec:data3}). If we only consider the infrared colors
(Fig.~\ref{fig:cmd2}a), the data spread of $lm$ and $pm$
stars is very significant though their mean values approximately
follow the MS position given by Koornneef (1983). It is
obvious that this spread is due to the infrared magnitude errors
in the 2MASS catalogue at the level of $K \approx 14-15$ as it is
strongly reduced when combined with optical data to obtain the
$V-K$ index (Fig.~\ref{fig:cmd2}b). So, the distance modulus fit
in the infrared diagrams turns out to be quite acceptable; in
addition, infrared diagrams independently confirm that the
reddening law is normal as suggested by the optical TCDs of
Fig.~\ref{fig:ccd1}.

As a final note, the infrared CMD of the $CF$ (see
Sect.~\ref{sec:member}) which is shown in Fig.~\ref{fig:cmd2}c
confirms that the region around the MS has almost no stars
reinforcing the real nature of the cluster.

\section{The Be-type star LSS 440} \label{sec:lss}

\begin{figure*}
\centering
\centerline{\psfig{file=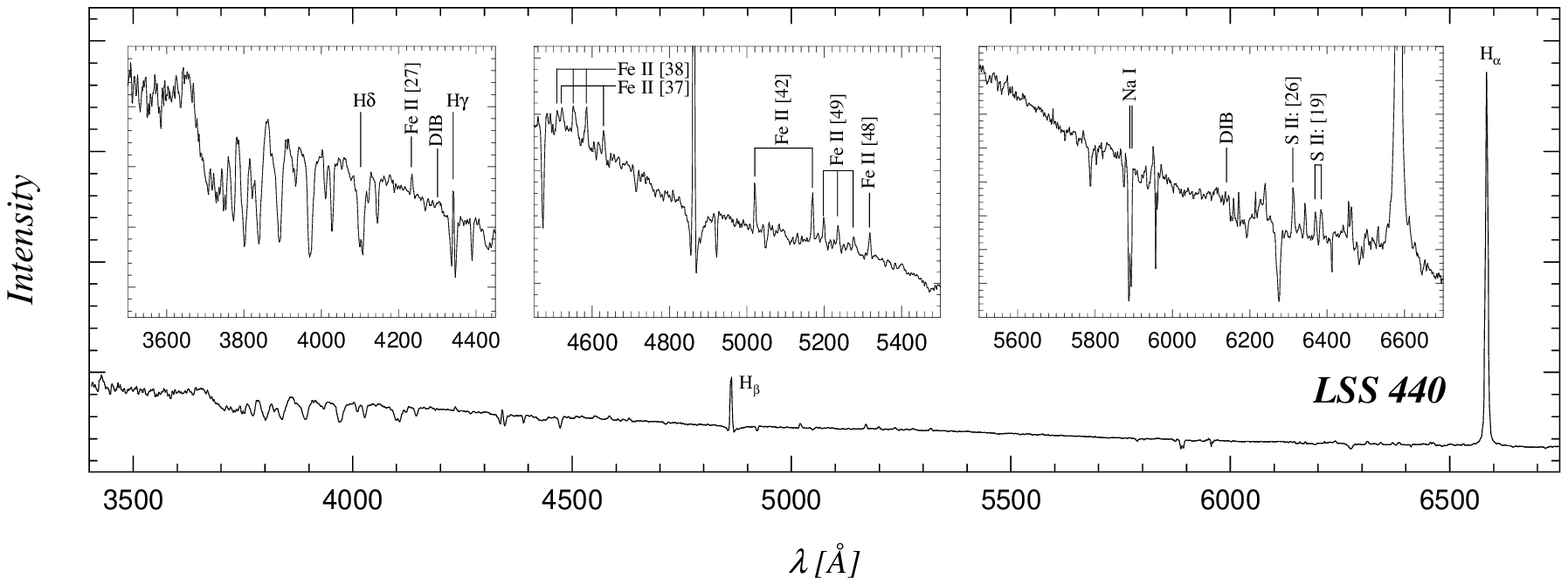,width=17cm}}
\caption{Complete observed spectrum of LSS 440 star on the main panel and its
main features in detail on the three upper minor panels}
\label{fig:lss440}
\end{figure*}

A singular object in the field of NGC~2401 is the bright star
$\#~14$ located at $1\farcm8$ south-east from the cluster
center. This star was early identified by Stephenson \& Sanduleak
(1971) as LSS~440 who informed it is an OB-type star with the
Balmer continuum in emission (in an exceptionally pronounced way)
and the $H_{\alpha}$ line in emission too according to an
independent $H_{\alpha}$ plate. To classify this star a series of
spectra were obtained covering the spectral range 3400 \AA~- 6750 \AA
(see Sect.~\ref{sec:data2}). The LSS~440 spectrum is shown
in Fig.~\ref{fig:lss440} together with a detail of the most
relevant features. This star shows the $H_{\alpha}$ line
in strong emission, together with $H_{\beta}$ and $H_{\gamma}$
lines in clear core emission and $H_{\delta}$ only in a weak way.
The Fe{\sc ii} 27, 37, 38, 42, 48, 49 and S{\sc ii} 19, 26
multiplets also appear in emission. The Balmer continuum is in
emission too, but not as strong as Stephenson \& Sanduleak (1971)
claimed. The $H_{\alpha}$ equivalent width is $\sim~76$ \AA~and
the rotation velocity is near $v~sin~i \sim 270-300$ km/sec from
He{\sc i} lines 4471 and 4026.

The spectral classification of LSS 440 was performed using
the BCD spectrophotometric system (see Barbier \& Chalonge 1941,
Chalonge \& Divan 1952, 1973 and Cidale et al. 2001 for the
description of the method). Briefly, the method is based on the
study of the Balmer discontinuity, which is independent on
interstellar and circumstellar extinctions. Using then the
obtained BCD parameters $\lambda_1 = 59.7$ \AA, $D = 0.078$ dex
and calibration tables given by Zorec (1986), we classified this
star as B0 Ve. The early type and the presence of emission FeII
lines could indicate LSS 440 as a member of Group I according to
the classification scheme proposed by Jaschek et al. (1980).

The models given by Zorec et al. (2002) can be used to get
the average photospheric properties for this star as well: $M_V =
-3.6 \pm 0.5$; $T_{eff} = 30000 \pm 1000~K$; $log~g = 4.0 \pm 0.1$
and $M_{bol} = -6.5 \pm 0.2$. Combining with Girardi et al. (2000)
evolutionary models of solar metallicity, a star with such
properties should have a mass near 15 ${\cal M}_{\sun}$
and an age about 5-6 $10^6$ yr.

Likewise, based on the slope change of the star spectra
from 4000 \AA~to 4600 \AA~, and using the Chalonge \& Divan (1973)
calibration improvements to the BCD method by Cidale et al.
(2001), we can compute the star visual absorption ($A_V$) produced
by both the interstellar dust and the circumstellar envelope. The
computations yielded $A_V = 1.33$ setting the true distance
modulus of LSS 440 in $V_O - M_V = 14.44 \pm 0.5$ corresponding to
a distance $d=7.7$ kpc.

\begin{figure}
\centering
\centerline{\psfig{file=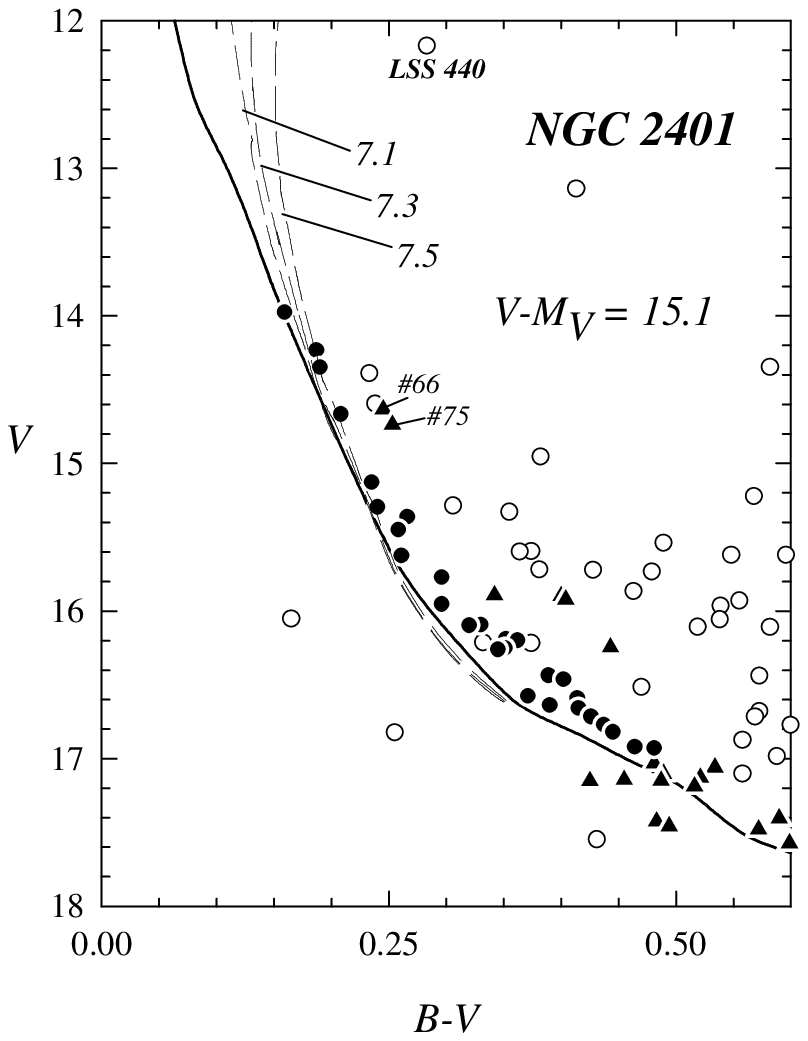,width=6cm}}
\caption{$V$ vs. $B-V$ CMD showing the isochrones (dashed lines) from Girardi
et al. (2000) and the Schmidt-Kaler (1982) ZAMS (solid line) corrected for the
effects of reddening and distance (see Sect.~\ref{sec:parameters}). Symbols have
the same meaning as in Fig.~\ref{fig:ccd1}. The numbers indicate $\log(age)$.}
\label{fig:age}
\end{figure}

Fabregat \& Torrejon (2000) studied the abundance of Be stars
in open clusters as a function of the cluster ages. Following
these authors findings Be-type stars show a maximum of appearance
in clusters the age of NGC 2401. From this point of view, to find
a Be-type star in this cluster should not be odd. However, the
distance of LSS 440 disagrees with the cluster distance and it is
placed a bit far from the cluster center. Likewise, the mass of
LSS 440 is extremely high if compared to the mass of next bright
member star onto the cluster MS (see Sect.~\ref{sec:parameters}
and Sect.~\ref{sec:LFIMF} in advance) which is $\approx 2 mag$
fainter. Certainly, the location of LSS 440 onto the $M_V$ vs.
$B-V$ and $U-B$ vs. $B-V$ diagrams are in agreement with the usual
location of other Be stars in open clusters (see Figs.~6a and 7
given by Mermilliod 1982) but the available elements at the moment
preclude any kind of clear physical relationship between the
cluster and this star. Radial velocities would be a very useful
tool to settle this question.

\section{Cluster luminosity and initial mass functions} \label{sec:LFIMF}

For the construction of the cluster luminosity function,
defined as the distribution of stars over the magnitude range in
bins $1^m$ wide, we applied the procedure already described in
Baume et al. (2004). That is, we first computed the apparent
magnitude distribution of $lm$ and $pm$ stars for $V < 17$ (LSS
440 star was not included in this analysis) and from the
substraction method for $V \geq 17$.

The latter procedure consists in a ``cleaning'' of the CMD of all
stars inside the cluster limits by removing the contribution of
field stars projected onto the cluster itself. We assume that our
$CF$ provides a good estimation of the contamination by field
interlopers and is valid across the $'cluster~area'$. We tried to
reduce the contamination coming from the Galactic disk population
by means of two envelope curves around the MS on the $V$ vs. $V-I$
plane (dashed curves on Figs.~\ref{fig:cmd1}ef). We located them
reasonably far from the MS in order to include all cluster stars.
Then we used a separation from the MS of about $0.3^m$ in color for a
given $V$ magnitude, a value that was increased to faint magnitudes
according to photometric errors. It is to be mentioned that the
completeness of our data has been estimated as in Baume et al. (2004)
and like in that case incompleteness is only severe at very faint
magnitudes ($V > 21$).

Once the entire distribution of apparent magnitudes is ready, it is
transformed into the $M_V$ distribution using the cluster distance modulus
of Sect.~\ref{sec:parameters}. The results are presented in Table~\ref{tab:lf}
in a self explanatory format.

\setcounter{table}{1}
\begin{table}
\caption[]{Stellar counts, completeness values and apparent cluster LF}
\label{tab:lf}
\begin{center}
\begin{tabular}{ccccc}
\hline \hline
$\Delta~V$ & Cluster & Comparison & Completeness & Apparent \\
           & Area    & Field      & [\%]         & LF       \\
\hline
14-15 & ~~~8 & ~2 & 100.0 & ~6 \\
15-16 & ~~20 & ~7 & 100.0 & 10 \\
16-17 & ~~38 & 11 & ~99.8 & 18 \\
17-18 & ~~55 & 42 & ~97.3 & 15 \\
18-19 & ~~94 & 59 & ~96.2 & 36 \\
19-20 &  106 & 69 & ~95.5 & 39 \\
20-21 & ~~99 & 85 & ~94.2 & 15 \\
21-22 &  110 & 98 & ~91.8 & 13 \\
22-23 &  125 & 99 & ~78.4 & 33 \\
23-24 & ~~46 & 19 & ~57.3 & 47 \\
\hline \hline
\end{tabular}
\end{center}
\end{table}

\begin{figure}
\centering
\centerline{\psfig{file=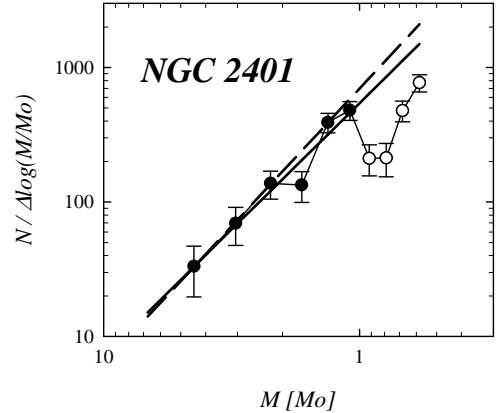,width=6.5cm}}
\caption{Initial Mass Function (IMF) of NGC~2401. Error bars are from Poisson
  statistics. The weighted least square fittings for the more massive bins are
  indicated by solid and dashed right lines (open symbols indicate bins not used
  in the fits. See Sect.~\ref{sec:LFIMF} for details).}
\label{fig:imf}
\end{figure}

Regarding the IMF, which is defined as the distribution of
original ZAMS stellar masses in logarithmic bins, absolute
magnitude intervals of the LF were converted into mass intervals
using the mass-luminosity relation given by Scalo (1986). What we
obtained is indeed the Present Day Mass Function (PDMF) which in
the case of NGC~2401 is very close to the IMF because only stars
below the ``turn-off'' are considered. The results shown in
Fig.~\ref{fig:imf} indicate that the IMF displays a constant slope
for the most massive bins (filled circles) while a small dip
appears at about $2~M_{\sun}$ followed by another one, more
noticeable, near $0.8~M_{\sun}$. Since the last (less massive)
mass-points (open circles) are dubious because of probable
incompleteness, the dip could be an artifact; if not, they are
suggesting a pronounced flattening of the mass function. The
mass-points of the most massive stars were fitted through a
weighted least squares fit that yielded a slope $x = 2.0 \pm 0.1$
for $M > 2~M_{\sun}$ and $x = 1.8 \pm 0.2$ for $M > 1~M_{\sun}$
(see Fig.~\ref{fig:imf}).

\section{A probable PMS star population in NGC 2401} \label{sec:PMS}

No doubt NGC~2401 is young and could still show hints of a
PMS star population which is normally placed over
the fainter part of the cluster above the MS. In fact,
Figs.~\ref{fig:cmd1}de suggest the presence of a ``turn--on''
point at $V \sim 17-18$; however the detection of a PMS star population
at this magnitude level is hard due to the field contamination.
Figs.~\ref{fig:cmd1}ab, on the other hand, are not deep enough to
perform a trustable analysis of this type.

In order to clarify this point, we applied again the substraction procedure
using Figs.~\ref{fig:cmd1}ef (see Sect.~\ref{sec:LFIMF}), but now
selecting also stars by $V-I$ bins $0.25^m$ wide (see Baume et al. 2003). This
way we could build an array indicating the amount of stars in each $V$ and $V-I$
box, both for the $'cluster~area'$ and for the $CF$. By performing then the difference
between these two arrays, we notice an excess of stars present approximately in
the path indicated by the long dashed right lines on Fig.~\ref{fig:cmd1}e. This
fact reinforces therefore the probable existence of a PMS star population in NGC~2401.
However, deeper photometric observations and/or spectroscopy of selected stars
in this region would be important to confirm this issue.

\section{NGC 2401 and the galactic structure in the Third Galactic Quadrant} \label{sec:disc}

The distance and age of NGC~2401 (see
Sect.~\ref{sec:parameters}) indicate this object belongs to the
innermost side young population of the probable extension of the
Norma-Cygnus spiral--arm in the Third Galactic Quadrant. This
outer spiral structure was already suggested by Vogt (1976) using
Luminous Stars and, more recently, by Kaltcheva \& Hildilch (2000)
studying OB stars. Star forming regions were found by Russeil
(2003), though an inspection of his Fig.~5 shows few young objects
at the galactic position of NGC~2401. The Georgelin \& Georgelin
(1976) list shows just one H$\alpha$ source, S~298, at a distance
of 5.2~kpc but with a galactic longitude of $227.7^{\circ}$. Also,
few open clusters with ages less than 100~Myr are found in this
region when inspecting the Dias et al. (2002) catalogue. It is
obvious that the Norma-Cygnus arm can be optically detectable
partially through windows in the interstellar absorption. However,
May et al. (2005) could trace this spiral--arm by means of CO
clouds and optical evidences of its existence have also been
recently reported by Carraro et al. (2005b) who detected the
presence of a large population of blue distant stars behind
several open clusters in this Galactic Quadrant. In this sense,
the cluster upper MS (see Fig.~\ref{fig:cmd1}) of NGC~2401 coincides
with the 'blue plume' seen in the cluster sample analyzed by Carraro
et al. (2005b) which they associate to a young population defining
the Norma-Cygnus spiral--arm. In fact, the brightest stars of
NGC~2401 appear at $V~\approx 14-15$ and $B-V~ \approx 0.1$ and
merge with old disc population at $V~\approx 18$ just as stars in the
'blue plume' of Carraro et al. (2005b) do. So, NGC~2401 is then part
of the small sample of very young open clusters presently known to
trace the Norma-Cygnus arm: Bochum~2, Haffner~18, Dolidze~25 and Pismis~1,
among others. The spatial distribution of CO clouds (May et al. 2005), the
young distant population (Carraro et al. 2005b) and the distance of
NGC~2401 suggest altogether it is located in the innermost side of
the Norma-Cygnus arm. Therefore, NGC~2401 CMDs can be used in the
future as a possible template to recognize and match features such
as the mentioned 'blue plume' seen in the optical CMDs of other
places in the Third Galactic Quadrant.

We also have shown arguments that weaken the probable
relationship of LSS 440 to NGC~2401: LSS 440 is placed at a
distance $d = 7.7$ kpc, but the error in its $M_V$ estimate and in
the cluster distance modulus may locate the star as far from the
Sun as 9.5 kpc or as close to it as 6.1 kpc. Only in this late
marginal case, some relation can be stated, but we have already
shown other arguments such as the star mass against that. By
ignoring any connection between the cluster and the star, it
becomes evident that unlike NGC 2401, LSS 440 is located well
inside the Norma-Cygnus spiral--arm so that, in any case, we are
dealing with very young and remote objects that reinforce the
spiral structure of the Milky Way in this quadrant of the Galaxy.

\section{Conclusions} \label{sec:conc}

We have presented a detailed multicolor photometric study
in the region of the open cluster NGC~2401, we confirmed the emission
nature of LSS 440 and gave its first spectral type classification.
In our opinion, the present state of knowledge of this star precludes
discard its membership to the cluster but no firm argument favoring its
cluster membership has been found. NGC~2401 is a very young object
($\sim 25$ Myr) placed at $6.3 \pm 0.5$~kpc from the Sun what makes
it an object located in the outskirts of our galaxy and therefore a
good tracer of the continuation of the Norma-Cygnus arm into the Third
Galactic Quadrant. Not confirmed at all, we found weak evidences of a
probable PMS star population accompanying this cluster. As for the
cluster IMF, the slope value found for NGC~2401 is not far from
the typical values shown by Scalo (1998, 2005).

\section*{acknowledgements}
  The authors thank the CASLEO staff for the technical support and the very
  useful discussions and valuable comments from L. Cidale. AM thanks the
  financial support from FCT (Portugal) grants BPD/20193/99, SFRH/BPD/19105/2004
  and the YALO project (PESO/P/PRO/1128/96). GB acknowledges a postdoctoral grant
  from Padova University where part of this work has been made. GC work has been
  partially supported by {\it  Fundaci\'on Andes}. This research has been also
  carried out under the cooperative international agreement Argentino-Italiano
  SECYT-MAE (IT/PA03 - UIII/077). We are much obliged for the use of the NASA
  Astrophysics Data System, of the Simbad database (Centre de Donn\'es Stellaires ---
  Strasbourg, France) and of the WEBDA open cluster database. This publication
  also made use of data from the Two Micron All Sky Survey, which is a joint
  project of the University of Massachusetts and the Infrared Processing and
  Analysis Center/California Institute of Technology, funded by the National
  Aeronautics and Space Administration and the National Science Foundation.

\end{document}